\documentclass{article}
\usepackage{spconf,amsmath,graphicx,amssymb,amsfonts,pifont}
\usepackage{url}
\usepackage[T1]{fontenc}
\usepackage{multirow}
\usepackage{multicol}
\usepackage{booktabs}
\usepackage{enumitem}

\usepackage{xcolor}

\setlist{leftmargin=3.5mm}

\makeatletter
\renewcommand\section{\@startsection {section}{1}{\z@}%
                                    {-2.0ex \@plus -0.5ex \@minus -0.5ex}%
                                    {1.0ex \@plus 0.5ex}%
                                    {\normalfont\Large\bfseries}}
\renewcommand\subsection{\@startsection{subsection}{2}{\z@}%
                                      {-1.0ex\@plus -0.25ex \@minus -.0ex}%
                                      {0.8ex \@plus 0.12ex}%
                                      {\normalfont\large\bfseries}}
\renewcommand\subsubsection{\@startsection{subsubsection}{3}{\z@}%
                                          {-0.25ex\@plus -0.1ex \@minus -.2ex}%
                                          {0.4ex \@plus .12ex}%
                                          {\normalfont\normalsize\bfseries}}
\renewcommand\paragraph{\@startsection{paragraph}{5}{\z@}%
                                      {0.5ex \@plus0.2ex \@minus1.01ex}%
                                      {-0.25em}%
                                      {\normalfont\normalsize\bfseries\slshape}}
\renewcommand\subparagraph{\@startsection{subparagraph}{6}{\parindent}%
                                         {0.25ex \@plus0.2ex \@minus .2ex}%
                                         {-0.2em}%
                                         {\normalfont\normalsize}}
\makeatother

\newcommand{\texttoken}[1]{{\small\textsf{#1}}}

\title{Resource-Efficient Adaptation of Speech Foundation Models \\for Multi-Speaker ASR}
\name{Weiqing Wang$^*$, Kunal Dhawan$^*$, Taejin Park, Krishna C. Puvvada, Ivan Medennikov, Somshubra Majumdar, He Huang, Jagadeesh Balam, Boris Ginsburg}
\address{NVIDIA, Santa Clara, USA}
\email{\{weiqingw,kdhawan,taejinp,kpuvvada,imedennikov,smajumdar,heh,jbalam,bginsburg\}@nvidia.com}
\begin{document}
%
\maketitle

\footnotetext[1]{The starred(*) authors contributed equally to this work.} 
\begin{abstract}
Speech foundation models have achieved state-of-the-art (SoTA) performance across various tasks, such as automatic speech recognition (ASR) in hundreds of languages. However, multi-speaker ASR remains a challenging task for these models due to data scarcity and sparsity. In this paper, we present approaches to enable speech foundation models to process and understand multi-speaker speech with limited training data. Specifically, we adapt a speech foundation model for the multi-speaker ASR task using only telephonic data. Remarkably, the adapted model also performs well on meeting data without any fine-tuning, demonstrating the generalization ability of our approach. We conduct several ablation studies to analyze the impact of different parameters and strategies on model performance. Our findings highlight the effectiveness of our methods. Results show that less parameters give better overall cpWER, which, although counter-intuitive, provides insights into adapting speech foundation models for multi-speaker ASR tasks with minimal annotated data.

\end{abstract}
\section{Introduction}
\label{sec:intro}
The field of automatic speech recognition (ASR) has recently shifted towards \textit{Foundation Models}, inspired by advancements in large language model (LLM) research. These models are typically very large, with over a billion parameters, and are trained on millions of hours of supervised multilingual speech data. The most notable speech foundation models in the literature currently are Whisper~\cite{radford2023robust}, Canary~\cite{puvvada2024less}, USM~\cite{zhang2023google}, and OWSM~\cite{peng2024owsm}.

Unlike traditional ASR tasks, speaker diarization and multi-speaker ASR tasks face significant challenges due to the lack of annotated real data. Consequently, most end-to-end frameworks rely on simulated data (also referred to as \textit{audio mixture})~\cite{kanda2020serialized, kanda2022streaming, cornell2024one, fujita2019end, horiguchi2020end} for training. However, reliance on simulated datasets introduces additional complexities. While these datasets can mimic real-world conditions to some extent, they often lack the variability and unpredictability found in natural conversations. This discrepancy can result in models that perform well on simulated data but struggle with real-world scenarios. Therefore, addressing data scarcity and sparsity problems is essential for the development of robust multi-speaker ASR systems. In addition to efforts in large-scale data collection, it is crucial to explore approaches for tasks with limited data.

In this paper, we present approaches to enable speech foundation models to process and understand multi-speaker speech, addressing the data scarcity problem. Specifically, we adapt a speech foundation model to the multi-speaker ASR task. Without altering the architecture or training objectives, we can extend a speech foundation model to this task using only limited data. Compared to state-of-the-art multi-speaker ASR models~\cite{cornell2024one, kanda2022transcribe}, our model is fully end-to-end and does not require clustering for speaker diarization. The key contributions of this work are:

\vspace{-0.3ex}
\begin{itemize}
\setlength\itemsep{0.05em}
\item \textbf{Adapter:}\; We successfully adapt the speech foundation model to the multi-speaker ASR task.
\item \textbf{Trade-off on scale:}\; Ablation studies show the trade-off between word error and speaker error. Our results show that using smaller adapters results in better overall cpWER, which is a counter-intuitive but interesting finding. 
\item \textbf{Generalizability:}\; We demonstrate that the speech foundation model with adapters shows better generalization across different domains. All models are trained on the same dataset, yet the adapters yield better performance in a different domain.

\end{itemize}

\section{Related Works}
\label{sec:related}
\subsection{Speaker diarization with ASR}

Before the concept of multi-speaker ASR became popular in the speech recognition community, the problem of \textit{“who spoke when”} was referred to as \textit{speaker diarization}. Speaker diarization involves labeling frame-level speaker assignments but does not include the task of transcribing speech to text (STT). Although there has been a clear distinction between speaker diarization and ASR, many studies have investigated the integration of speaker diarization modules with ASR.

The earliest approach to using lexical information from ASR output for speaker diarization was the RT03 evaluation~\cite{tranter2003investigation}, which utilized word boundary information for segmentation. Despite the modest performance gain, this was the first attempt to enhance diarization with ASR-based segmentation. Subsequent systems, such as those described in~\cite{huang2007ibm} and~\cite{silovsky2012incorporation}, refined speaker activity detection (SAD) and addressed word-breakage issues, respectively. Additionally,~\cite{canseco2004speaker} developed a dictionary for common broadcast news phrases to identify speakers, demonstrating the evolution of integrating ASR output to improve diarization.

More recent speaker diarization systems use neural models to capture linguistic patterns from ASR output, enhancing diarization results. The system in~\cite{flemotomos2019linguistically} significantly improved diarization error rate (DER) by combining linguistic and acoustic information using a neural text-based speaker change detector and role recognizer. Additionally,~\cite{park2018multimodal} and~\cite{park2019speaker} utilized lexical information for speaker segmentation and clustering, demonstrating that integrating lexical and acoustic information improves diarization performance.

When it comes to modular (pipelined) multi-speaker ASR, the CHiME Challenges~\cite{vincent20164th, watanabe2020chime, cornell2023chime} have greatly advanced the field. The multi-speaker ASR system proposed in~\cite{medennikov2020stc} for the CHiME-6 Challenge~\cite{watanabe2020chime} gained significant attention. In this system, speaker-specific masks are estimated via a target-speaker voice activity detector (TS-VAD) to transcribe multi-channel, multi-speaker audio mixtures.

\subsection{End-to-end multi-speaker ASR}

The multi-speaker ASR system proposed in~\cite{qian2018single} demonstrated that an end-to-end multi-speaker ASR system can be built using multiple single-speaker ASR modules. The concept of recognizing multiple speakers using a single ASR model was introduced in~\cite{yu17b_interspeech}, which focused on two speakers and employed permutation-free training with RNN-T-based ASR modules. A method integrating source separation and speech recognition into a single system was proposed in~\cite{seki2018end}, utilizing CTC-based loss for the ASR components. Another approach, presented in~\cite{shafey2019joint}, introduced a jointly-trainable RNN-T-based two-speaker ASR model that generates speaker labels and text tokens simultaneously. The system in~\cite{chang2019end} employed attention mechanisms to train an end-to-end multi-speaker ASR model without using alignments or non-mixture labels for each speaker. This was further extended to a multi-channel concept in~\cite{chang2019mimo}, resulting in an end-to-end Multi-channel Input Multiple Output (MIMO) system.

One of the most significant advancements in end-to-end multi-speaker ASR was the introduction of Serialized Output Training (SOT) in~\cite{kanda2020serialized}. SOT utilizes attention mechanisms similar to~\cite{chang2019end, chang2019mimo}, but it does not require multiple encoders or multiple heads. Instead, SOT leverages the multi-head attention mechanism in Transformer architecture~\cite{vaswani2017attention}. This approach eliminates the need for neural mask estimators, which are commonly used in other studies such as~\cite{chang2019mimo, vzmolikova2019speakerbeam, boeddeker2018front}. The multi-head attention-based ASR system focuses on speakers through activations in the Feed-Forward Network (FFN) of the Transformer architecture, allowing it to handle overlapping speech by utilizing various attention values from multiple attention heads. This makes the multi-speaker ASR system simpler by relying solely on the multi-head self-attention mechanism. The SOT approach was later extended to t-SOT (token-level SOT) for streaming systems~\cite{kanda2022streaming}. Recently, an end-to-end joint speaker diarization and speech recognition system was proposed in~\cite{cornell2024one}, which performs multi-speaker ASR and clusters the speakers later in the process.

\section{Multi-speaker Adaptation}
In this section, we first introduce the foundation model and then explain the adaptation for the proposed multi-speaker ASR system. 

\subsection{Speech foundation model}

We utilize the pretrained Canary-1B~\cite{nvidia2024canary} as the foundation model. Canary-1B is an open-source Automatic Speech Recognition (ASR) and Automatic Speech Translation (AST) model supporting English, French, Spanish, and German. Similar to Whisper, this model incorporates specific prompt tokens, including \texttoken{<|startoftranscript|>}, \texttoken{<|transcribe|>}, \texttoken{<|translate|>}, \texttoken{<|nospeech|>}, \texttoken{<|endoftranscript|>}, and an additional special token for each supported language. Furthermore, Canary-1B employs SentencePiece~\cite{kudo2018sentencepiece} and a concatenated tokenizer~\cite{dhawan2023unified} with a vocabulary size of 1024 for each language, alongside a 32-unit sub-tokenizer for the special tokens.

\subsection{Dataset preparation}
\label{DatasetPreparation}

\begin{figure}[!t]
\centering
\includegraphics[width=0.48\textwidth]{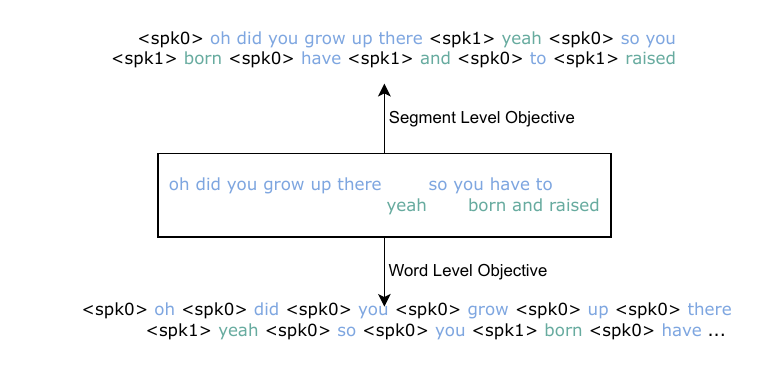}
\caption{How to generate the segment and word level objective. }
\label{fig:SOT}
\end{figure}

\begin{figure*}[ht]
\centering
\includegraphics[width=0.8\textwidth]{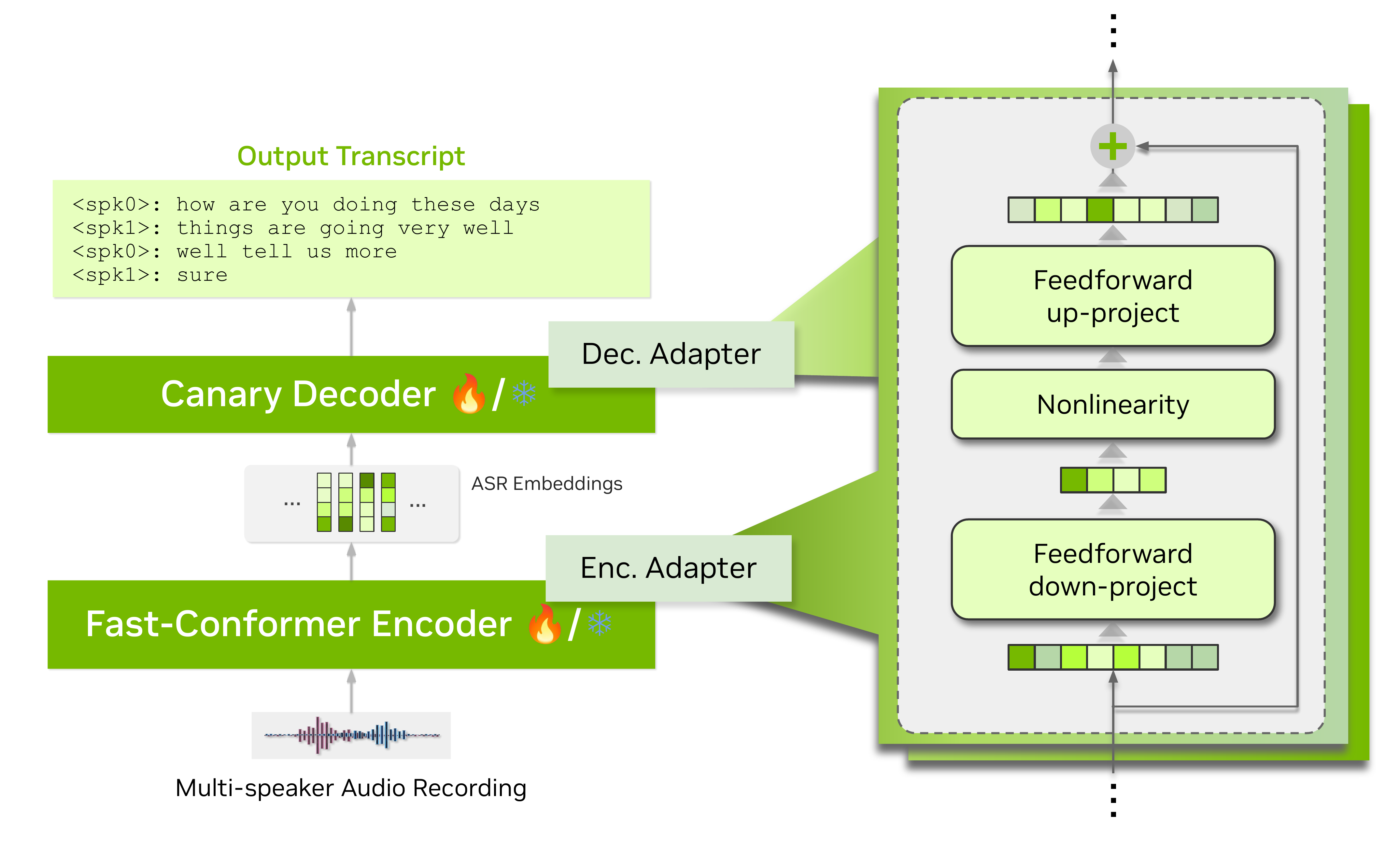}
\caption{Canary Encoder-Decoder Structures and Adapters for Multi-speaker ASR}
\label{fig:msasr_adapter}
\end{figure*}

Similar to the SOT~\cite{kanda2020serialized} and t-SOT~\cite{kanda2022streaming} frameworks, we introduce special tokens into the original transcripts for the multi-speaker ASR task. However, unlike these frameworks, we insert a speaker token at each speaker change instead of a speaker change token. For example, the multi-speaker transcripts always start with \texttoken{<|spk0|>}, followed by the sentence from that speaker. When a speaker change occurs, we insert \texttoken{<|spk1|>} for the new speaker. We employ a first-in-first-out (FIFO) speaker assignment strategy: \texttoken{<|spk0|>} for the first speaker, \texttoken{<|spk1|>} for the second speaker, and so on. This approach is similar to that in~\cite{cornell2024one}.

Handling overlapping speech is challengeing for speaker diarization and multi-speaker ASR. The simplest method assigns the speaker token based on the start time of each word, treating overlapping speech as a frequently changing speaker region. 
In this way we can generate the segment level objective for training.

A major issue with segment level objective is that if a speaker token is missed in the predicted transcript, subsequent tokens may be assigned to the wrong speaker, leading to significant errors. This is similar to some cascade speaker diarization systems, where the segments are generated by speaker change detection~\cite{lu2002speaker}. To address this, we try to generate a word level objective for training. As shown in Fig. \ref{fig:SOT}, we insert a speaker token for each word, resulting in extremely choppy transcripts.

\subsection{Adaptation for multi-speaker ASR}

Unlike the single-speaker ASR task where million hours of data is available, we usually do not have enough labeled data to train a robust multi-speaker ASR model. In addition, the pretrained speech foundation model will forget previous tasks when fully fine-tuned on unseen tasks~\cite{houlsby2019parameter}. Specifically, the only difference between the single-speaker ASR and the multi-speaker ASR is that we just insert speaker tokens into the single-speaker transcripts. In that case, the previous learned knowledge is more important for the multi-speaker ASR. Therefore, adapter~\cite{houlsby2019parameter} becomes a better choice for this task, where only a small number of additional, learnable parameters are introduced to the frozen foundation model. Here, the foundation model can still keep the the previously learned ASR capabilities, and adapters focus more on the speaker information.

To keep the foundation model frozen and only train the adapter, we also keep the tokenizer unchanged. The unused special tokens in the special tokenizer are employed as the speaker tokens. 

Fig. \ref{fig:msasr_adapter} shows the architecture of the Canary-1B model and the adapters. We apply an adapter module to the output of each conformer/transformer layer in the encoder and decoder. Each adapter consists of two linear layers with an activation function in between. The adapter also includes a residual connection, allowing it to approximate an identity function with near-zero initialization~\cite{houlsby2019parameter}.

\newcommand{\cmark}{\ding{51}}%
\newcommand{\xmark}{\ding   {55}}%
\begin{table*}[htbp!]
  \caption{ Evaluation results on the Callhome 2-speaker sessions. `\xmark' stands for frozen encoder/decoder and `\cmark' stands for unfrozen encoder/decoder. }
  \label{tab:results_callhome}
  \centering
  \vspace{2ex}
  \begin{tabular}[c]{cccccccccc}
    \toprule
     \multirow{2}*{\textbf{Adapter}} & \multirow{2}*{\textbf{Num Trainable }} & \multirow{2}*{\textbf{Encoder}} & \multirow{2}*{\textbf{Decoder}} & \multicolumn{3}{c}{\textbf{Segment-level Obj.}} & \multicolumn{3}{c}{\textbf{Word-level Obj.}}  \\
    \cmidrule(lr){5-7} \cmidrule(lr){8-10}
     \textbf{Dimension} & \textbf{Parameters} & & & WER & cpWER & $\Delta$cp & WER & cpWER & $\Delta$cp\\
     
   \midrule
        32 & 3.2 M & \xmark & \xmark & \textbf{16.90} & \textbf{20.40} & 3.50 & 19.24 & 24.83 & 5.59 \\
	64 & 6.4 M & \xmark & \xmark & 17.92 & 21.44 & 3.52 & \textbf{19.12} & \textbf{22.63} & 3.51 \\
	128 & 12.7 M & \xmark & \xmark & 19.08 & 22.27 & 3.19 & 19.82 & 23.11 & 3.29 \\
	256 & 25.3 M & \xmark & \xmark & 19.71 & 21.85 & \textbf{2.14} & 19.81 & 22.68 & \textbf{2.87} \\
        \midrule \multirow{3}*{-}
	 & - & \xmark & \xmark & 20.02 & - & - & - & - & - \\
	 & 609 M & \xmark & \cmark & 23.30 & 30.28 & 6.98 & 23.03 & 31.75 & 8.72 \\
	 & 1.0 B& \cmark & \cmark & 21.70 & 25.07 & 3.37 & 21.82 & 24.59 & 2.77 \\
    \midrule
        \multicolumn{4}{c}{Baseline} & 23.77 & 26.67 & 2.90 & - & - & -  \\
     \bottomrule
     \end{tabular}
\end{table*}

\section{Experimental Setup}

\subsection{Dataset}

\subsubsection{Speech Foundation Model}
Canary-1B is trained on a multilingual dataset comprising English, German, Spanish, and French, totaling 81.5K hours of speech. More details can be found in~\cite{puvvada2024less}.

\subsubsection{Multi-speaker ASR}
For the multi-speaker ASR task, we use the Fisher dataset~\cite{cieri2004fisher} for training, which contains 2k hours of speech. Since the Fisher dataset includes only two speakers, we evaluate our model solely on the 2-speaker sessions in the Callhome~\cite{callhome} and AMI~\cite{carletta2005ami} datasets.

Given that the Fisher dataset provides single-channel speech for each speaker, we perform force alignment on the single-speaker speech to obtain word-level timestamps and generate SOT-style transcripts as described in Section \ref{DatasetPreparation}, including both the segment-level objective and word-level objective. Each session is divided into 10 to 20-second segments for training and evaluation.

The Fisher dataset labels are not always accurate, with some words occasionally missing. To address this, we clean the data by removing segments with high word error rates (WER). Specifically, we use Canary-1B\footnote{\url{https://huggingface.co/nvidia/canary-1b}} and Parakeet-ctc-1.1B\footnote{\url{https://huggingface.co/nvidia/parakeet-ctc-1.1b}} to transcribe each segment and eliminate those with a WER exceeding 80\%. 

\subsection{Evaluation metric}
We evaluated the model using word error rate (WER), concatenated minimum-permutation word error rate (cpWER)~\cite{watanabe2020chime}, and $\Delta$cp~\cite{park2024enhancing}. The cpWER is computed as follows: (i) concatenate all utterances of each speaker for both the reference and hypothesis files; (ii) compute the WER between the reference and all possible speaker permutations of the hypothesis; (iii) select the lowest WER among them, which is assumed to be the best permutation. The cpWER includes errors from both speech recognition and speaker diarization. Additionally, we assume that the difference between cpWER and WER reflects the diarization error. Therefore, we use $\Delta$cp = cpWER - WER as a metric for our evaluation.

\subsection{Model}
\subsubsection{Speech Foundation Model}
The Canary-1B model uses a FastConformer encoder~\cite{rekesh2023fast} and a Transformer decoder, containing one billion parameters in total. FastConformer is a speech-specific modification of the transformer based on Conformer~\cite{gulati2020conformer} that increases the downsampling factor to 8, achieving a 2.8x speedup without loss of modeling capacity~\cite{rekesh2023fast}. Both the encoder and decoder have 24 layers.

\subsubsection{Adapters}
As shown in Fig. \ref{fig:msasr_adapter}, we utilize a straightforward architecture for the adapters: two linear layers with an activation function in between. Each adapter layer is applied directly after each Conformer/Transformer layer. We gradually increase the dimension of the linear layers to validate the performance with different numbers of parameters.

\subsection{Training details}
The model is trained on the Fisher dataset and evaluated on the 2-speaker sessions of the Callhome and AMI datasets. Five percent of the Fisher dataset is used as the validation set. The audio signals are segmented into 10-20 seconds and utilize 80-dimensional log Mel-filterbank energies with a frame length of 25 ms and a frame shift of 10 ms for acoustic features. Additionally, data augmentation is performed using background noise from Musan~\cite{snyder2015musan}.

During the training phase, both the encoder and decoder are frozen, and only the adapters are trained. The model is trained for 50K steps using the NVIDIA NeMo~\cite{kuchaiev2019nemo} framework. The AdamW optimizer is employed with a peak learning rate of 3e-4. The learning rate is warmed up over 2.5K steps and annealed according to the Noam scheduling. Lhotse~\cite{zelasko2021lhotse} is used for data loading with a batch duration of 360 seconds per GPU and a quadratic duration of 15 seconds. Each model was trained using an NVIDIA RTX A6000 (48GB) GPU for about 12 hours.

\begin{table*}[htbp!]
  \caption{ Evaluation results on the AMI 2-speaker sessions. `\xmark' stands for frozen encoder/decoder and `\cmark' stands for unfrozen encoder/decoder. The models are ONLY trained with Fisher and have never seen AMI data. }
  \label{tab:results_ami}
  \centering
  \vspace{2ex}
  \begin{tabular}[c]{cccccccccc}
    \toprule
     \multirow{2}*{\textbf{Adapter}} & \multirow{2}*{\textbf{ Num Trainable }} & \multirow{2}*{\textbf{Encoder}} & \multirow{2}*{\textbf{Decoder}} & \multicolumn{3}{c}{\textbf{Segment-level Obj.}} & \multicolumn{3}{c}{\textbf{Word-level Obj.}}  \\
    \cmidrule(lr){5-7} \cmidrule(lr){8-10}
     \textbf{Dimension} & \textbf{Parameters} & & & WER & cpWER & $\Delta$cp & WER & cpWER & $\Delta$cp\\
   \midrule
        32 & 3.2 M & \xmark & \xmark & \textbf{24.13} & \textbf{27.81} & 3.68 & \textbf{25.15} & 29.34 & 4.19 \\
	64 & 6.4 M & \xmark & \xmark & 25.79 & 29.43 & 3.64 & 25.52 & \textbf{29.05} & \textbf{3.53} \\
	  128 & 12.7 M & \xmark & \xmark & 27.16 & 30.82 & 3.66 & 25.92 & 29.59 & 3.67 \\
	  256 & 25.3 M & \xmark & \xmark & 28.24 & 30.85 & \textbf{2.61} & 25.75 & 29.99 & 4.24 \\
        \midrule \multirow{3}*{-}
	& - & \xmark & \xmark & 22.34 & - & - & - & - & - \\
	& 609 M & \xmark & \cmark & 28.78 & 37.50 & 8.72 & 29.28 & 39.44 & 10.16 \\
	& 1 B & \cmark & \cmark & 31.14 & 35.78 & 4.64 & 31.35 & 35.99 & 4.64 \\
    \midrule
        \multicolumn{4}{c}{Baseline} & 24.05 & 34.22 & 10.17 & - & - & -  \\
     \bottomrule
     \end{tabular}
\end{table*}

\section{Experimental Results}

We conducted several ablation experiments for this task, examining the impact of various factors such as the number of parameters, different transcripts, and cross-domain evaluations. The model was trained ONLY with the Fisher dataset and evaluated on the Callhome and AMI datasets. We trained the model with both segment-level and word-level objectives, as previously mentioned in Fig. \ref{fig:SOT}.

\subsection{Baseline}
The baseline results were obtained from the clustering-based speaker diarization with ASR\footnote{\url{https://github.com/NVIDIA/NeMo/blob/main/examples/speaker_tasks/diarization/clustering_diarizer/offline_diar_with_asr_infer.py}} from the NeMo toolkit. For the Fisher dataset, we followed the telephonic configuration, and for the AMI dataset, we followed the meeting configuration.

\subsection{Number of parameters}
For the adapters, we kept the entire foundation model frozen and only trained the adapters. We gradually increased the size of the linear layers and the number of parameters of the adapters. Additionally, we directly fine-tuned the speech foundation model without adapters on the multi-speaker ASR task. In this context, we can either partially fine-tune the decoder while keeping the encoder frozen, or fully fine-tune the entire model. If both the encoder and decoder are frozen, the model will not produce any speaker tokens and will treat it as a single-speaker ASR task; hence, we only provide the WER for this model.

Table \ref{tab:results_callhome} shows the results for different numbers of parameters. For the segment-level objective, as we increase the size of the adapters, the WER increases, but the $\Delta$cp decreases. Adapters with more parameters tend to learn more speaker information for the speech foundation model, which may confuse the previously learned ASR knowledge, leading to a lower $\Delta$cp but higher WER. Similar trends can be observed for the word-level objective.

Table \ref{tab:results_callhome} also shows the results when we directly fine-tune the model on the multi-speaker ASR task. If we freeze the encoder and only train the decoder, the WER is close to that of the model with adapters, but the $\Delta$cp is worse. This is because the speech foundation model has not learned any speaker-related tasks, and the higher layers contain more text-related information. Even when fine-tuning the decoder, it is difficult to recognize speaker identity with text-only information. Once we unfreeze the encoder, the performance improves significantly, but it is still worse than the model with adapters.

One significant difference between adapter training and full fine-tuning is that using adapters does not degrade ASR performance and even shows lower WER than the speech foundation model. However, the fully fine-tuned model exhibits worse ASR performance, suggesting that adapters are a better choice for the multi-speaker ASR task.

\subsection{Segment-wise objective v.s. word-wise objective}

We evaluated the model using both segment-level and word-level transcripts to determine the effectiveness of each approach. As shown in Table \ref{tab:results_callhome}, both segment-level and word-level objectives produced comparable results in terms of cpWER, but WER does not improve as the number of adapter parameters is reduced.

The segment-level objective tends to produce lower WER values across most configurations compared to the word-level objective. For instance, with a 32-dimensional adapter, the WER is 16.90\% for the segment-level objective, while the WER is 19.24\% for the word-level objective. Both objectives show a similar trend where increasing the adapter dimensions initially improves $\Delta$cp but does not consistently improve WER. This suggests that while the model becomes better at handling speaker error (as indicated by $\Delta$cp), the overall WER does not necessarily benefit from larger adapter dimensions. The $\Delta$cp values for the word-level objective are generally higher than those for the segment-level objective, indicating that the segment-level approach better handles speaker transitions.



     

\subsection{Cross-domain evaluation}

As mentioned in previous sections, the model is only trained and evaluated on telephonic data. We also evaluate this model on the AMI dataset, which is in a different domain—meetings. Note that we do not fine-tune or train any model on the AMI dataset. As Table \ref{tab:results_ami} shows, the model trained with adapters performs better than the fine-tuned foundation model. This indicates that adapters also provide better generalization ability for unseen domains.

Comparatively, the Canary-1B model achieved a segment-level WER of 22.34\%, which is close to the performance of the adapter-trained model, indicating that the adapter approach does not significantly degrade performance in a new domain. However, the models fine-tuned on the foundation model showed significantly worse results. The model with a frozen encoder and trainable decoder had a WER of 28.78\% and cpWER of 37.50\% when trained with the segment-level objective. The fully fine-tuned model performed even worse, with a WER of 31.14\% and cpWER of 35.78\% with the segment-level objective.

These comparisons underscore the superior generalization ability of models trained with adapters. Despite being trained solely on telephonic data, the adapter-trained models maintained robust performance when evaluated on the AMI meeting dataset, significantly outperforming the fully fine-tuned foundation models. This suggests that using adapters not only preserves the model’s capability to generalize across different domains but also enhances it, making adapters a more effective solution for cross-domain multi-speaker ASR tasks.

\section{Conclusion}
In this paper, we addressed the challenge of multi-speaker ASR with data scarcity and sparsity using adapters. By adapting a speech foundation model with only telephonic data, we demonstrated significant improvements in processing and understanding multi-speaker speech. Our adapted model not only performed well on telephonic data but also showed remarkable generalization to meeting data without any fine-tuning on domain-specific data. Through various ablation studies, we analyzed the impact of different parameters and strategies on model performance, revealing the trade-offs between word error and speaker error. Our findings highlight the effectiveness of our adaptation methods and provide valuable insights into optimizing speech foundation models for multi-speaker ASR tasks with minimal annotated data. This work underscores the potential for extending speech foundation models to new tasks with limited data, paving the way for more robust and versatile ASR systems. In the future, we plan to extend this framework to streaming scenarios with more speakers.


\bibliographystyle{IEEEbib}
\bibliography{refs}

\end{document}